\documentclass[aps,pre,twocolumn,superscriptaddress,showpacs]{revtex4-1}

\usepackage{amssymb}
\usepackage{color}
\usepackage{graphicx}
\usepackage{amsmath}
\usepackage{mathrsfs}
\usepackage{times}
\usepackage{subeqnarray}
\usepackage{cases}
\usepackage{bm}

\begin{document}

\title{Enhancing network synchronization by phase modulation}

\author{Huawei Fan}
\affiliation{School of Physics and Information Technology, Shaanxi Normal University, Xi'an 710062, China}

\author{Ying-Cheng Lai}
\affiliation{School of Physics and Information Technology, Shaanxi Normal University, Xi'an 710062, China}
\affiliation{School of Electrical, Computer, and Energy Engineering, Arizona State University, Tempe, Arizona 85287, USA}

\author{Shi-Xian Qu}
\affiliation{School of Physics and Information Technology, Shaanxi Normal University, Xi'an 710062, China}

\author{Xingang Wang}
\email[Email address:]{wangxg@snnu.edu.cn}
\affiliation{School of Physics and Information Technology, Shaanxi Normal University, Xi'an 710062, China}

\begin{abstract}

Due to time delays in signal transmission and processing, phase lags are 
inevitable in realistic complex oscillator networks. Conventional wisdom is
that phase lags are detrimental to network synchronization. Here we show that 
judiciously chosen phase lag modulations can result in significantly enhanced 
network synchronization. We justify our strategy of phase modulation, 
demonstrate its power in facilitating and enhancing network synchronization 
with synthetic and empirical network models, and provide an analytic 
understanding of the underlying mechanism. Our work provides a new approach 
to synchronization optimization in complex networks, with insights into 
control of complex nonlinear networks.

\end{abstract}

\date{\today}
\maketitle

\section{Introduction} \label{sec:intro}

The function and operation of realistic complex systems rely on the 
coherent motion of their constituent dynamical elements, generating a 
continuous interest in the synchronization behaviors of coupled 
oscillators~\cite{SYNBOOK:Kuramoto,SYNBOOK:Pikovsky,SYNREV:Boccaletti,
DSN:RMP,SYNREV:Arenas}. In the study of oscillator synchronization, an issue of theoretical and practical significance 
is how to achieve global synchronization in a large-scale networked system 
at reduced coupling cost~\cite{SYNREV:Boccaletti,DSN:RMP,SYNREV:Arenas}.
In the past, a number of strategies were proposed and studied to address 
the problem of synchronization optimization in coupled oscillator networks 
systems~\cite{AEM:2005,NT:2006,XGW:2007,HL:2008,ZA:2011,SUNJ}. For example, 
the small-world and scale-free features in engineering and natural 
systems~\cite{SWNET,SFN:1999} have been exploited to achieve optimal 
synchronization in complex networks~\cite{PLM:2002,YM:2004,OA,
XGW:Chaos,GGJ,LMH:2011}. 

A paradigm in the study of network synchronization is the Kuramoto 
model~\cite{SYNBOOK:Kuramoto}, in which an ensemble of phase oscillators 
with distributed natural frequencies are coupled in a nonlinear fashion 
and can be locked in phase when the mutual coupling parameter exceeds 
a critical value. The Kuramoto model and its generalized forms capture 
the essence of the collective dynamics of many realistic 
systems~\cite{JAA:RMP,RFA:2016,BS:ExpSyn,WS:1993}, and have been exploited
as a standard model for testifying the efficiency of various synchronization 
optimization strategies~\cite{SUNJ,SWNET,YM:2004,OA,XGW:Chaos,GGJ,LMH:2011,
ZZG:1998,WUYE,OEO:2012,DSP:2013,ZHL,BHJ:2016}. In terms of optimization, 
the existing synchronization strategies can be classified into two major 
categories: one based on adjusting the network structure and another 
seeking/arranging optimal locations for oscillators. Strategies based on 
network structural perturbations are applicable to situations where the 
nodal dynamics are fixed but there is flexibility in modifying the link 
structure such as the addition of shortcuts~\cite{SWNET,YM:2004}. The 
alternative strategy applies to situations where the network structure cannot 
be altered but it is possible to rearrange the locations of the oscillators 
according to the network structure~\cite{SUNJ,GGJ,WUYE}. There have been 
theoretical efforts in establishing the efficiency of the various 
synchronization optimization strategies. However, to physically realize  
certain strategies, e.g., to supply shortcut links in a real neuronal 
network or to relocate a pair of power stations in an actual power grid,
may be difficult. It remains to be an open and challenging problem to 
articulate physically reasonable strategies to optimize synchronization in 
complex networks of coupled nonlinear oscillators.

Due to the limited speed of signal transmission and processing, time delay 
is ubiquitous in real world systems. For a network of coupled oscillators, 
a time delay represents a phase lag in the interaction. In previous studies,
time delays are generally considered as detrimental to 
synchronization~\cite{OEO:2012,DSP:2013,SK:1986,MAL:2015,EM:2011,DP:2011,
chimera_Kuramoto,chimera_Abrams,WSL:2009}. When a uniform phase lag exists
in the coupling, global synchronization can be destroyed, leading to the
emergence of alternative types of collective dynamics on the network.
For example, when the value of the phase lag is about $\pi/2$, a chimera 
state~\cite{chimera_Kuramoto,chimera_Abrams} can arise, in which both 
synchronized and unsynchronized groups of oscillators coexist. For $\pi$ phase
lag, anti-phase synchronization clusters can form~\cite{TLS:2005,LMH:2010}. 
For random phase lags, the onset of synchronization can be significantly 
delayed or even disappear~\cite{MAL:2015,EM:2011}. Motivated by the ubiquity
of time delay and its strong ability to alter the network synchronization
dynamics, we ask the following question of both basic and practical 
significance: is it possible to exploit phase lag modulations for 
synchronization optimization? More specifically, we ask whether it would
be possible to apply judiciously chosen phase lags to enhance network
synchronization, as in applications where synchronous dynamics are desired. 
The main contribution of this paper is an affirmative answer to this question.

In general, the amount of phase lag for an oscillator will depend on the 
its intrinsic dynamics or state, and the phase lags cannot be expected to be uniform
among the oscillators in the network. An advantage of this approach is that
phase modulation does not require the adjustment of the network structure or 
relocation of the oscillators. Our phase modulation based strategy thus holds 
the promise of achieving optimal, highly efficient synchronization in complex
oscillator networks at low cost. More specifically, we consider complex 
networks of coupled phase oscillators with fixed network structure and 
oscillator arrangement. We introduce to each oscillator a constant phase lag 
whose amount is determined by the natural frequency of the oscillator. We 
demonstrate numerically and argue analytically that the proposed phase 
modulation scheme is capable of dramatically improving network synchronization,
as characterized by a marked increase in the value of the synchronization 
order parameter and a significant decrease in the critical coupling strength 
for global synchronization. The phase modulation strategy is physically 
realizable and practically implementable, opening a new approach to 
investigating optimization and control of collective dynamics in complex 
nonlinear networks. 

In Sec.~\ref{sec:model}, we describe networked systems of coupled phase
oscillators, introduce the phase-modulation strategy, and provide numerical
results on the effect of phase modulation on synchronization. In 
Sec.~\ref{sec:theory}, we present a theoretical analysis to explain the 
underlying mechanism enhancing synchronization. In Sec.~\ref{sec:application},
we apply our phase modulation strategy to two empirical systems: the 
neuronal network of nematode C. Elegans and a power grid network. 
In Sec.~\ref{sec:partial}, we present results of optimizing synchronization
using a partial phase modulation scheme and compare the performance of the phase-modulation strategy to that of two recent
optimization approaches. In Sec.~\ref{sec:discussion}, we summarize and
present a discussion of the main result.

\section{Model and phenomena} \label{sec:model}

Our model of networked phase oscillators reads
\begin{equation} \label{eq:model}
\dot{\theta}_i=\omega_i+\frac{K}{d_i}\sum^{N}_{j=1}a_{ij}
\sin{(\theta_j-\theta_i-\alpha_i)}, 
\end{equation}
where $i,j=1,\ldots,N$ are the oscillator (node) indices, $\theta_i(t)$ is
the instant phase of the $i$th oscillator at time $t$, and $K$ is the 
uniform coupling strength. The natural frequency of the $i$th oscillator is 
$\omega_i$, which follows the distribution $g(\omega)$. The coupling 
relationship of the oscillators is described by the adjacency matrix 
$A=\{a_{ij}\}$, with $a_{ij}=a_{ji}=1$ if oscillators $i$ and $j$ are 
directly coupled by a link, otherwise $a_{ij}=0$. The quantity 
$d_i=\sum_j a_{ij}$ is the number of connections associated to oscillator $i$, 
i.e., its degree, and $\alpha_i$ is the phase modulation introduced to 
oscillator $i$, which depends on $\omega_i$ (with a specific form given 
below). Equation~(\ref{eq:model}) is similar in form to the generalized 
Kuramoto-Sakaguchi model, which has been extensively studied in the 
literature for various synchronization phenomena among networked 
oscillators~\cite{SK:1986,MAL:2015,EM:2011,DP:2011,chimera_Kuramoto,
chimera_Abrams,TLS:2005,LMH:2010}.   

Different from existing models where $\alpha_i$ is uniform or randomly
distributed, we judiciously set $\alpha_i$ based on the information of 
the oscillator dynamics. The specific form of $\alpha_i$ and the reason 
behind are the following. For an ensemble of coupled phase oscillators 
whose natural frequencies follow a given distribution, the occurrence of 
synchronization is determined by two counter-balancing factors: coupling 
and frequency spread~\cite{JAA:RMP,RFA:2016,BS:ExpSyn}. Whereas the former 
tends to make the oscillators coherent, the latter prevents this tendency. 
In the presence of coupling, the effective frequency of each oscillator is 
different from its natural frequency. As the coupling strength is increased 
from zero, the spread of the effective frequencies will be gradually 
narrowed. At the critical point of global synchronization, the effective 
frequencies of the oscillators become identical, giving rise to the 
synchronous motion. This intuitive picture gives an important indication 
on how to enhance synchronization by modulating phase lags in 
Eq.~(\ref{eq:model}): narrowing the distribution of the effective 
frequencies of the oscillators by tuning $\alpha_i$.   

Assume that the system is in the vicinity of the global synchronization 
state: $\theta_{i}(t)\thickapprox\theta_{j}(t)$ for $i,j=1,\ldots,N$. 
Equation~(\ref{eq:model}) can be approximated as
\begin{equation}
\dot{\theta}_i=\tilde{\omega}_i +\frac{K}{d_i}\sum^{N}_{j=1}a_{ij}
(\theta_j-\theta_i)\cos{(\alpha_i)},
\end{equation}
where $\tilde{\omega}_i=\omega_i-K\sin{\alpha_i}$ is the modified natural 
frequency of oscillator $i$. Requiring $\tilde{\omega}_i=0$, we have 
$\omega_i-K\sin{\alpha_i}=0$, which gives $\alpha_i=\arcsin(\omega_i/K)$. 
Setting phase lags this way, the modified natural frequencies of all the 
oscillators will be identical, facilitating synchronization. To keep 
$\tilde{\omega}_i=0$ for all the oscillators, it is necessary that 
$K\ge K_c\equiv\omega_{max}$ be satisfied, with 
$\omega_{max}=\max\{|\omega_i|\}$ being the largest natural frequency of 
the oscillators. To make $\alpha_i$ independent of $K$ (so that $\alpha_i$ 
is a constant value for each oscillator), we introduce the phase modulation
scheme
\begin{equation} \label{eq:lag}
\alpha_i=\arcsin(\beta\omega_i),
\end{equation}
with $\beta\in[0,\beta_{max}]$ being the modulation amplitude. Here the 
quantity $\beta_{max}=1/K_c=1/\omega_{max}$ is the largest modulation 
amplitude capable of generating identical effective frequency. With 
Eq.~(\ref{eq:lag}), the modified natural frequencies can be rewritten as 
$\tilde{\omega}_i = \omega_i (1-K\beta)$. 

The effects of phase modulation on synchronization can be intuitively 
understood, as follows. For a fixed value of $\beta$, as $K$ is increased 
from zero, the distribution of $\tilde{\omega}_i$ will be gradually narrowed, 
resulting in enhanced synchronization. At the critical point 
$K^{\beta}_c=1/\beta$, we have $\tilde{\omega}_i=0$ for all oscillators, 
signifying that global synchronization has been achieved. 
For $K > K^{\beta}_c$, $\tilde{\omega}_i$ spreads out again, deteriorating 
synchronization. As $\beta$ is increased from zero to $\beta_{max}$, the 
critical coupling strength $K^{\beta}_c$ will decrease gradually. For 
$\beta=\beta_{max}$, we get the minimum critical coupling strength for 
global synchronization: $K_c=1/\beta_{max}=\omega_{max}$.

\begin{figure*}
\centering
\includegraphics[width=0.8\linewidth]{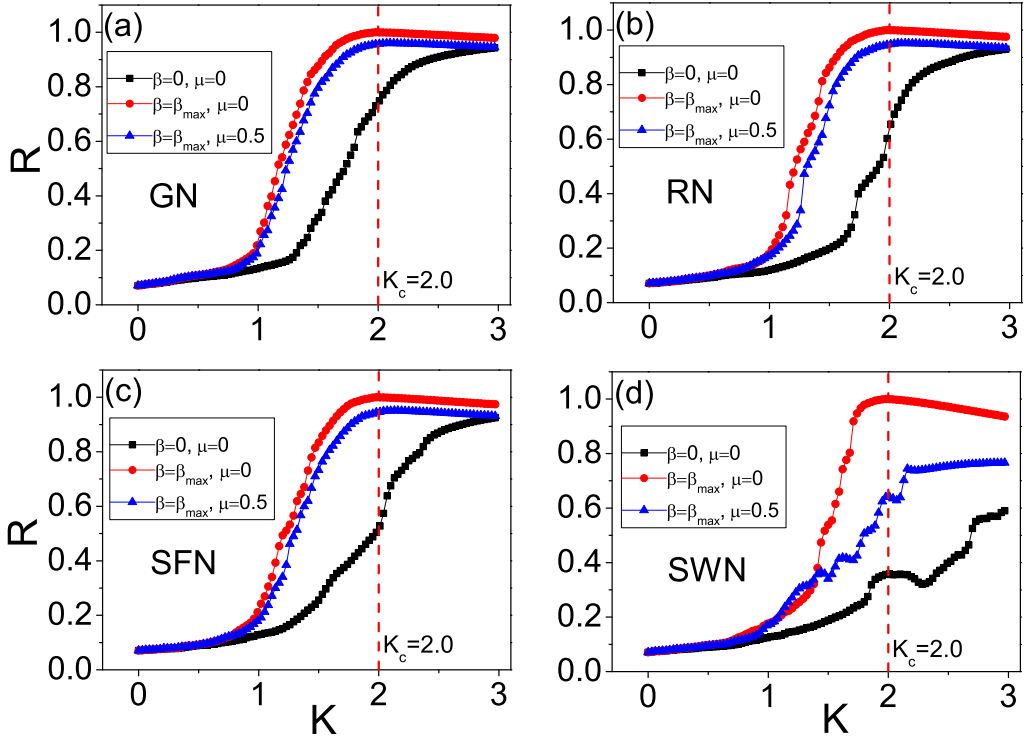}
\caption{(Color online) The impact of phase modulation on synchronization 
in different network models. (a-d) For globally connected (GN), random (RN), 
scale-free (SFN), and small-world (SWN) networks, the order parameter $R$ 
versus the coupling strength $K$ with or without phase modulation and 
noise perturbations: $(\beta,\mu)=(0,0)$ (black squares), 
$(\beta,\mu)=(\beta_{max},\mu)=(0.5,0)$ (red circles), 
$(\beta,\mu)=(0.5,0.5)$ (blue triangles). The network size is $N=200$. 
The average degree of the networks in (b-d) is $\langle k\rangle=16$. The 
rewiring probability for the small-world networks is $0.1$. In all network 
models, the critical value of the coupling parameter is 
$K_c = 1/\beta_{max}=\omega_{max}=2.0$, which agrees with the theoretical 
prediction. Each data point is the result of averaging over $20$ 
statistical realizations.} 
\label{fig1}
\end{figure*}

To test the proposed phase modulation scheme, we carry out simulations
using a variety of network models. In all cases, the size of the network 
is $N=200$ and the natural frequencies of the oscillators follow the 
truncated Lorentzian distribution:
$g(\omega) = (\delta/\pi)/[(\omega-\omega_0)^{2}+\delta^2]$.
The central frequency, scale parameter and truncation frequency of the 
Lorentzian distribution are $\omega_0=0$, $\delta=1$, and $\omega_{max}=2.0$, 
respectively. Network synchronization is characterized by the order parameter 
\begin{displaymath}
R=\langle |\sum_{j=1}^N e^{i\theta_j}|\rangle
\end{displaymath}
where $|\cdot|$ and $\langle\cdot\rangle$ denote, respectively, the module 
and time averaged functions, and $R\in[0,1]$, with $R=0$ and $1$ corresponding
to completely incoherent and global synchronization states, respectively. 
Figures~\ref{fig1}(a-d) show, for globally connected, random~\cite{ER}, 
scale-free~\cite{SFN:1999}, and small-world networks~\cite{SWNET}, 
respectively, the order parameter $R$ versus the coupling parameter $K$ for 
different values of the modulation amplitude $\beta$. We see that, 
when phase modulation is applied, the values of the order parameters 
are consistently larger than those without phase modulation, and transition 
to global synchronization characterized by $R = 1$ occurs for much smaller 
value of $K$. It is noted that, with phase modulation, the
critical value of the coupling parameter at which $R$ reaches unity is 
$K_c=\omega_{max}=2.0$, regardless of the network structure. (This property is rooted in the normalized coupling strategy we have adopted in the model.) As $K$ is increased further, the value of $R$ tends to reduce slightly. 

To assess the robustness of the phase modulation strategy in facilitating
and enhancing network synchronization, we perturb the phase lags by 
introducing independent and identically distributed random noise: 
$\alpha_{i}=\alpha_{0i}+\mu W_{i}$, where $\alpha_{0i}$ is the phase 
lag given by Eq.~(\ref{eq:lag}), $W_{i}$ is a random number uniformly 
distributed in the interval $[-1,1]$, and $\mu$ is the noise amplitude. 
Figure~\ref{fig1} shows that, while random noise tends to reduce the 
value of the order parameter and thus weaken synchronization, the negative
effect on synchronization is insignificant, suggesting the robustness
of the phase-modulation-based strategy to noise perturbations.  

\begin{figure}
\centering
\includegraphics[width=0.95\linewidth]{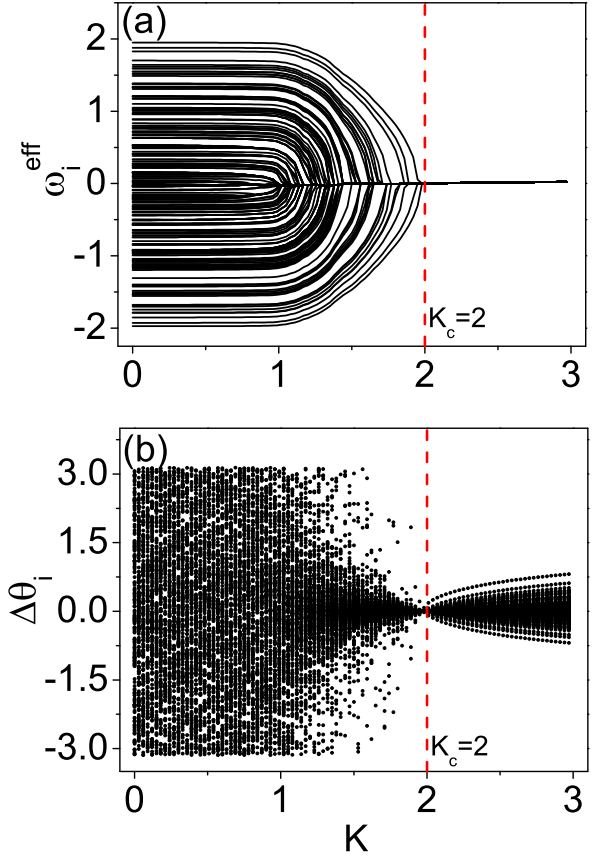}
\caption{ Behaviors of frequency and phase in the presence of phase 
modulation. For the model of globally connected network in Fig.~\ref{fig1}(a), 
the distribution of (a) the effective frequency $\omega^{eff}_i$ and (b) the 
phase difference $\Delta\theta$ taken at $t=1\times 10^3$ versus the coupling 
strength $K$. In (a), all the effective frequencies converge to zero for 
$K\ge K_c$. In (b), the phase differences first converge for $K<K_c$ and then 
diverge for $K>K_c$. At the critical point $K=K_c$, all the phase differences 
vanish.}   
\label{fig2}
\end{figure}

To gain further insights into the role of phase modulation in network 
synchronization, we examine the variations of the effective frequencies, 
$\omega^{eff}_{i}=(1/T)\int_{t}^{t+T}\dot{\theta}_{i}(\tau)d\tau$, with 
respect to $K$, where $\dot{\theta}_i(\tau)$ stands for the instantaneous 
frequency of oscillator $i$. Figure~\ref{fig2}(a) shows the numerical 
results for the model of globally connected network. We see that, as $K$
is increased from zero, the distribution of the effective frequencies is 
gradually narrowed. As the critical coupling strength $K_c$ is reached, 
all effective frequencies converge to a single value. Figure~\ref{fig2}(b) 
shows the variations of the phase difference, 
$\Delta\theta_{i}=\theta_{i}-\psi$, where $\psi$ is the average phase 
defined by the relation $e^{i\psi}=\sum_j e^{i\theta_j}/R$. We see that 
the phase differences are random for $K < K_c$, become zero at $K_c$, and 
exhibit a systematic delayed behavior as $K$ is increased through $K_c$. 
This phenomenon is consistent with the numerical results in 
Fig.~\ref{fig1}, where $R$ exhibits a slightly decreasing trend as the
value of $K$ is increased through $K_c$. For models other than the globally  
connected network, we obtain essentially the same results. 

The results in Figs.~\ref{fig1} and \ref{fig2} reveal the following 
physical picture about the transition to synchronization in the presence 
of phase modulation. As the value of $K$ is increased from zero, the 
difference between the modulated natural frequencies [i.e., $\tilde{\omega}_i$ 
in Eq.~(\ref{eq:lag})] gradually decreases, resulting in the convergence 
of the effective frequencies [Fig.~\ref{fig2}(a)] and an increase in the 
value of the order parameter [Fig.~\ref{fig1}]. At $K=K_c$, we have 
$\tilde{\omega}_i=0$ for all the oscillators and, due to the strong 
coupling strength, the phases of the oscillators are nearly identical, 
leading to the highest level of synchronization ($R\approx 1$). As the 
value of $K$ is increased from $K_c$, $\tilde{\omega}_i$ become 
non-identical again. However, since the value of $K$ is large, the 
oscillators are still locked in frequency [Fig.~\ref{fig2}(a)] but with 
scattered phases [Fig.~\ref{fig2}(b)]. It is the scattered phases which 
lead to the decrease of $R$ in the parameter region $K>K_c$ [Fig.~\ref{fig1}].

\section{Theoretical analysis} \label{sec:theory} 

We use the Ott-Antonsen ansatz~\cite{OA} to obtain an analytic understanding 
of the mechanism underlying phase-modulation enhanced synchronization. For 
the model of globally connected network [Figs.~\ref{fig1}(a) and \ref{fig2}], 
the dynamics of the phase oscillators can be rewritten as
\begin{eqnarray}
\dot{\theta}_i = \omega_i &+& [K/(2i)][(re^{-i\theta_{i}}-cc)\cos{\alpha_i} 
\nonumber \\
&-&i(re^{-i\theta_i} + cc)\sin{\alpha_i}],
\end{eqnarray} 
where $cc$ stands for the complex conjugate and $r=Re^{i\psi}$. In the 
thermodynamic limit $N\rightarrow\infty$, the state of the system at time 
$t$ can be described by a probability density function 
$f[\omega(\alpha),\theta,t] \equiv f(\omega,\theta,t)$,
whose evolution is governed by the continuity equation
\begin{eqnarray}
\partial f/\partial t &+& \partial\{[\omega+\frac{K}{2i}
(re^{-i\theta_{i}}-c.c.)h_{2}(\omega) \nonumber \\
&-&\frac{K}{2}(re^{-i\theta_{i}}+c.c.)h_{1}(\omega)]f\}/\partial\theta =0,
\end{eqnarray}
where $h_2(\omega)=\sqrt{1-\beta^2\omega^2}=\cos{\alpha} \in (0,1)$
and $h_1(\omega)=\beta\omega=\sin{\alpha} \in (-1,1)$. The complex order 
parameter can be expressed as
$r = \int_{0}^{2\pi}d\theta\int^{\infty}_{\infty}d\omega 
f(\omega,\theta,t)e^{i\theta}$. 

Expanding $f(\omega,\theta,t)$ as a Fourier series in $\theta$, we have
\begin{equation}\label{fs}
f=\frac{g(\omega)}{2\pi}[1+\sum^{\infty}_{n=1}f_{n}(\omega,t)e^{in\theta}
+\sum^{-1}_{n=-\infty}f^{*}_{n}(\omega,t)e^{in\theta}].
\end{equation}
Utilizing the Ott-Antonsen ansatz~\cite{OA}: 
$f_n(\omega,t)=a(\omega,t)^{n}$, for $|a(\omega,t)|\leq 1$,
we obtain
\begin{equation} \label{eq:aeq}
\frac{\partial a}{\partial t}+i\omega a+\frac{K}{2}(ra^{2}-r^{*})h_{2}
(\omega)-\frac{K}{2}(ira^{2}+ir^{*})h_{1}(\omega)=0
\end{equation}
and
\begin{equation}\label{rstar}
r^{*}=\int^{\infty}_{-\infty}d\omega a(\omega,t)g(\omega).
\end{equation}
We use the Lorentzian distribution for the natural frequencies: 
$g(\omega) =1/(i2\pi)[(\omega-\omega_0-i\delta)^{-1}
-(\omega-\omega_0+i\delta)^{-1}]$. 
To evaluate the integral that gives $r^*$, we analytically continue $\omega$ 
to the complex $\omega$-plane and carry out contour integration~\cite{OA},
where the analyticity of $\alpha$ holds in the lower half plane of the complex
variable $\omega$. For large negative values of $\mbox{Im}(\omega)$, 
Eq.~(\ref{eq:aeq}) can be approximated as 
$\partial a/\partial t=\mbox{Im}(\omega)a$, so we have 
$a\rightarrow 0$ for $\mbox{Im}(\omega)\rightarrow -\infty$. Following the
setting of numerical simulations, we have $\omega_{0}=0$ and $\delta=1$ for 
the Lorentzian distribution. The pole of the lower half plane is $\omega=-i$. 
We thus obtain $r=a^{*}(-i,t)$. Substituting this expression into 
Eq.~(\ref{eq:aeq}), we get the following nonlinear equation for the
order parameter
\begin{equation}\label{eq:rreq}
\frac{\partial R}{\partial t}+[1-\frac{K}{2}(\sqrt{1+\beta^{2}}+\beta)]R
+\frac{K}{2}(\sqrt{1+\beta^2}-\beta)R^3=0,
\end{equation}
which is the Bernoulli equation with a proper solution given by
\begin{equation} \label{eq:AS}
\frac{R(t)}{R_{0}}=\{1+(\frac{R_{0}^{2}}{R(0)^{2}}-1)
e^{2t[1-\frac{K}{2}(\sqrt{1+\beta^{2}}+\beta)]}\}^{-1/2},
\end{equation}
where 
\begin{displaymath}
R_{0}=|\frac{\sqrt{1+\beta^{2}}-\beta}{\sqrt{1+\beta^{2}}
+ \beta-\frac{2}{K}}|^{-\frac{1}{2}}. 
\end{displaymath}
We see that for $K<K_{0}=2/(\sqrt{1+\beta^{2}}+\beta)$, $R(t)$ tends to zero 
exponentially with time. For $K>K_{0}$, $R(t)$ approaches asymptotically a 
finite value $R_0$, where $K_0$ is the critical coupling at which the value 
of $R$ begins to increase from zero, i.e., onset of 
synchronization~\cite{XGW:Chaos,LMH:2011}. Setting $R_0=1$, we can obtain 
the critical coupling strength for global synchronization as 
$K^{\beta}_c=1/\beta$, which agrees with the heuristic argument in 
Sec.~\ref{sec:model}.

\begin{figure}
\centering
\includegraphics[width=\linewidth]{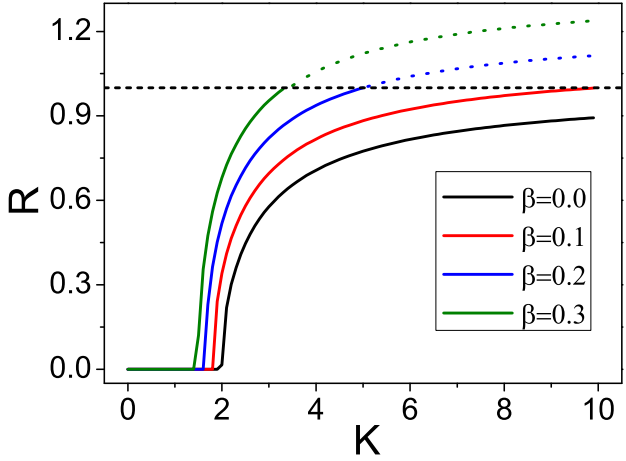}
\caption{(Color online) Theoretically predicted values of the order 
parameter in the presence of phase modulation. For the model of globally 
connected network, the order parameter $R$ calculated from the analytic 
prediction Eq.~(\ref{eq:AS}) versus the coupling parameter $K$ for different 
amplitudes of phase modulation: $\beta=0$ (black), $\beta=0.1$ (red), 
$\beta=0.2$ (blue), and $\beta=0.3$ (green). With phase modulation, $R$ 
reaches unity at a finite value of the coupling strength: $K_c=1/\beta$. 
However, without phase modulation, this occurs for $K_c\rightarrow \infty$ 
(black).} 
\label{fig3}
\end{figure}

Figure~\ref{fig3} shows the theoretically predicted value of $R$ from
Eq.~(\ref{eq:AS}) versus $K$ for different values of $\beta$. We observe
two features. Firstly, prior to the onset of global synchronization 
($K<K_c$), the value of the order parameter $R$ increases monotonically 
with $\beta$. Secondly, as the value of $\beta$ is increased from zero, 
the critical coupling strength for global synchronization gradually 
decreases. Specifically, Eq.~(\ref{eq:AS}) predicts that, without phase 
modulation ($\beta=0$), the onset of global synchronization occurs at 
infinite coupling strength ($K_c\rightarrow \infty$), whereas with 
phase modulation, this happens at a finite value: $K_c=1/\beta$. The 
theoretical prediction agrees well with the numerical results in 
Figs.~\ref{fig1} and \ref{fig2}. 

While Eq.~(\ref{eq:AS}) predicts that the $R$ values can be greater than 
unity, they are not physically meaningful. As depicted in Fig.~\ref{fig2}(b), 
once $K$ exceeds the critical value $K_c$, phase lags among the oscillators 
will set in, causing $R$ to decrease from unity. The seeming contradiction 
between the theoretical and numerical results for $K>K_c$ can be explained, 
as follows. Substituting $a=|a|e^{i\varphi}$ into Eq.~(\ref{eq:aeq}), we get
\begin{eqnarray}
\partial|a|/\partial t &+& (K/2)(|a|^{2}-1)\mbox{Re}
(re^{-i\varphi})h_{2}(\omega) \nonumber \\
&+& (K/2)(|a|^{2}+1)\mbox{Im}(re^{-i\varphi})h_{1}(\omega)=0,
\end{eqnarray}
which indicates that, quite different from the case without phase 
modulation~\cite{OA}, $\partial|a|/\partial t \neq 0$ for $|a|=1$. This 
means that a trajectory  of Eq.~(\ref{eq:aeq}) originated from some initial 
condition satisfying $|a|<1$ may cross the unit circle in the complex 
$a$-plane. Since $r=a^{*}(-i,t)$, the situation of $R>1$ can occur. In 
fact, for $|a|>1$, the Ott-Antonsen ansatz is no longer valid~\cite{OA}, 
making the analytic prediction for this regime physically meaningless. 
    
\section{Applications} \label{sec:application}

\begin{figure*}[tbp]
\centering
\includegraphics[width=0.8\linewidth]{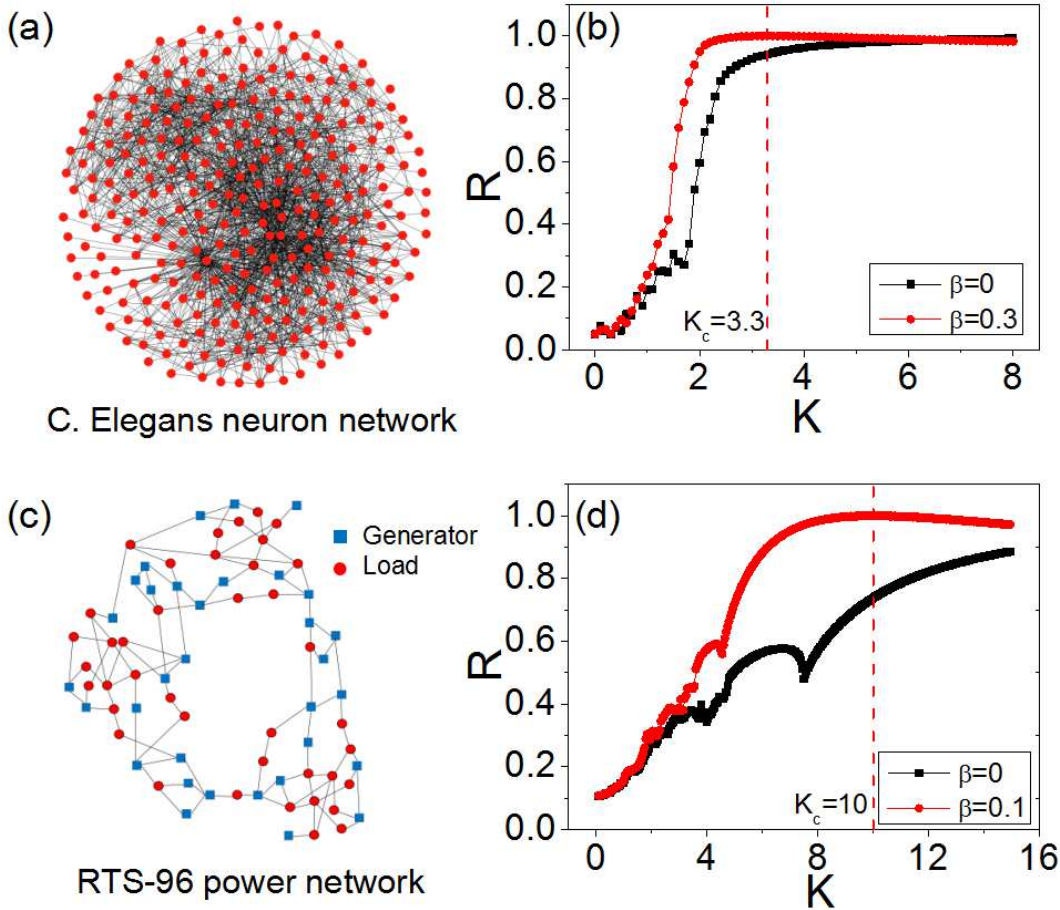}
\caption{(Color online) Synchronization optimization by phase modulation 
in realistic networks. (a) The structure of the neuronal network of the 
nematode C.~Elegans. (b) For the neuronal network, $R$ versus $K$ when 
phase modulation is absent ($\beta=0$, black squares) and present 
($\beta=0.3$, red circles). (c) The structure of the RTS-96 power grid. 
Blue squares and red nodes denote, respectively, generators and loads. 
(d) For the power-grid network, the variation of synchronization order 
parameter, $R$, with respect to the coupling strength, $K$, for the cases 
where phase modulation is absent ($\beta=0$, black squares) and present 
($\beta=0.1$, red circles). For both networks, the natural frequency 
of the nodes follows the Lorentz distribution, with the truncation 
frequency $\max\{|\omega|\}=2.0$, the location parameter $\omega_0=0$ and 
the scale parameter $\delta=1$. Results in (b) and (d) are averaged over 
$20$ frequency realizations.}  
\label{fig4}
\end{figure*}

We demonstrate the enhancement of synchronization by the strategy of 
phase modulation in two realistic systems. The first example is the 
neuronal network of the nematode C.~Elegans~\cite{WSTB:1986}, in which 
phase lags in coupling interaction can be modulated using methods 
such as drug delivery~\cite{BWK:2001}. The structure of the neuronal 
network is shown in Fig.~\ref{fig4}(a), which consists of $N = 297$ nodes 
and $2148$ links. Following Refs.~\cite{Varela:2001,Glass:2001}, we 
represent the dynamics of isolated neurons by non-identical phase 
oscillators and investigate their collective behavior using the generalized 
Kuramoto model Eq.~(\ref{eq:model}). We adopt the Lorentzian distribution 
for the natural frequencies and set $\alpha_i = \arcsin{(\beta \omega_i)}$. 
The numerical results with and without phase modulation are presented in 
Fig.~\ref{fig4}(b). We see that phase modulation can dramatically 
facilitate synchronization. Particularly, without phase modulation, 
global synchronization occurs at $K_c\approx 8$, whereas with phase 
modulation, this occurs at $K_c=1/\beta\approx 3.3$.

The second example is the RTS-$96$ power grid, whose structure is 
shown in Fig.~\ref{fig4}(c), which has $N=73$ nodes and $214$ links. 
To be realistic, we divide the nodes into two groups, generators and 
loads, and model their dynamics by the generic swing 
equations~\cite{PowerGrid:NT,DCB:2013}: 
\begin{eqnarray} \label{eq:PGM}
\nonumber
M_i\ddot{\theta}_i &+& D_i\dot{\theta}_i = 
\omega_i + \frac{K}{d_i}\sum_{j=1}^{N}a_{ij}
\sin{(\theta_{j}-\theta_{i}-\alpha_{i})}, \ i\in V_{1}, \\ 
D_l\dot{\theta}_l &=& \omega_l + \frac{K}{d_l}\sum_{j=1}^{N}a_{lj}
\sin{(\theta_{j}-\theta_{l}-\alpha_{l})}, \ l\in V_{2},
\end{eqnarray}
where $V_1$ and $V_2$ denote the sets of generator and load nodes, 
respectively, $M_i$ is the inertial coefficient for generator node $i$, 
and $D_i$ is the viscous damping coefficient. Adopting the Lorentzian 
frequency distribution and setting $\alpha_i = \arcsin{(\beta \omega_i)}$, 
we obtain the variation of $R$ versus $K$ with and without phase modulation,
as shown in Fig.~\ref{fig4}(d). (In simulations, we increase $K$ 
adiabatically from zero so as to avoid the problem of multiple coexisting 
attractors~\cite{PJM:2014}.) We see that, with phase modulation, 
synchronization is both facilitated and enhanced. In particular, without 
phase modulation, global synchronization is not reached even for $K=16$, 
whereas even with weak phase modulation ($\beta=0.1$), global 
synchronization occurs at $K_c\approx 10$. For a realistic power grid, 
phase modulation can be realized by adjusting the power angles between the
exciting and terminal voltages~\cite{VP:book}.

\section{Partial phase modulation scheme and comparison with two recent
optimization approaches} \label{sec:partial}

Our study has focused on the case where phase modulation is introduced to 
each oscillator in the network, this is not necessary in practical 
applications. In fact, network synchronization can still be significantly 
enhanced if phase modulation is introduced to only a fraction of the 
oscillators. To demonstrate this, we use the globally connected network 
in Fig.~\ref{fig1}(a) and introduce phase modulation to only $p=20\%$ of 
the oscillators, with the targeting oscillators selected in a descending 
order of the oscillator natural frequency. The numerical results are 
presented in Fig.~\ref{fig4}(a). We see that, comparing with the case 
without phase modulation, synchronization with partial phase modulation 
is significantly enhanced. As the fraction $p$ of modulated oscillators is 
increased, the synchronization performance approaches that with global 
phase modulation - the upper limit. 

\begin{figure}
\centering
\includegraphics[width=0.9\linewidth]{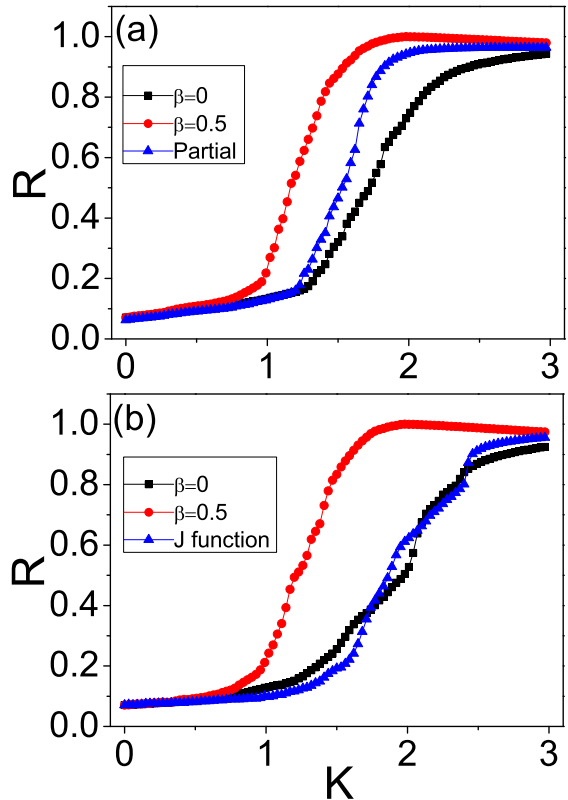}
\caption{(Color online) Effect of partial phase modulation. 
(a) For the model of globally connected network in Fig.~\ref{fig1}(a), 
order parameter $R$ versus the coupling strength $K$ for partial phase 
modulation (blue triangles) in which phase lags are introduced to 
only $p=10\%$ of the oscillators. (b) For the model of scale-free 
network in Fig.~\ref{fig1}(b), $R$ versus $K$ for phase modulation and 
J-function based synchronization strategies. In both panels, the results 
without ($\beta=0$, black squares) and with global ($\beta=0.5$, red circles) 
phase modulation are also included for comparison.}
\label{fig5}
\end{figure}

Our work is related to but distinct from two recent studies:
explosive synchronization~\cite{GGJ} and J-function-based synchronization 
optimization~\cite{SUNJ}. In explosive synchronization, it is the 
counter-balance between heterogeneous network structure and non-identical 
nodal dynamics which results in a first-order transition to 
synchronization~\cite{BS:ExpSyn}, whereas in our present work  
heterogeneity in the nodal dynamics is balanced by phase lags. 
Different from explosive synchronization, our phase modulation strategy 
does not require any resetting of the oscillator frequencies and, in terms 
of synchronization optimization, outperforms the frequency-weighted 
strategy adopted in explosive synchronization. For instance, for the model 
of globally connected network (with the same coupling scheme and frequency 
distribution used in our present work), the critical forward coupling 
strength in explosive synchronization is always larger than~\cite{ZHL} 
$K_c=2.0$, while through phase modulation the critical coupling strength 
for global synchronization is universally $K_c=2.0$ for all network 
models [Fig.~\ref{fig1}]. The J-function method, on the other hand, is 
a synchronization-optimization-oriented function~\cite{SUNJ} aiming to
improve synchronization in complex network of coupled phase oscillators. 
The J-function, which combines the information of network structure 
(i.e., the eigenvectors of the network coupling matrix) and oscillator 
dynamics (e.g., the natural frequencies of the oscillator), provides 
an efficient numerical approach to finding the optimal configuration 
of oscillators in large networks. Using the model of scale-free network 
in Fig.~\ref{fig1}(c), we compare the synchronization performance of 
the J-function-based and phase-modulation-based strategies. (In searching 
for the optimal configuration, we iterate the J-function $1\times 10^6$ 
times, following the method proposed in Ref.~\cite{SUNJ}. To make the 
comparison on an equal footing, we adopt the same ensemble of natural 
frequencies for the oscillators.) The results are presented in 
Fig.~\ref{fig4}. We see that the performance of the J-function-based 
strategy is lower than that of phase-modulation-based strategy. 
As the scheme of normalized coupling strength is concerned [Eq.~(2)], 
the optimal configuration has almost the same synchronization performance 
as that of random configuration, as shown in Fig.~\ref{fig4}(b).          

\section{Discussion} \label{sec:discussion}

To summarize, we have proposed a phase-modulation based strategy for 
facilitating and enhancing synchronization in complex networks and 
demonstrated its efficiency using a variety of network models as well as 
two realistic networks. The strategy should be distinguished from the 
existing methods of synchronization optimization in that it does not 
require any adjustment of the network structure or the locations of 
the oscillators. More importantly, our work demonstrates that time 
delays, in contrast to conventional wisdom that they are detrimental 
to synchronization, can actually benefit synchronization. As time delays are
ubiquitous in natural systems and phase lags can be implemented in 
practice, our strategy is physically realizable for realistic complex 
systems such as infrastructure and biological networks. Our study 
provides a novel approach to optimizing synchronization in complex 
networks of coupled nonlinear oscillators, with potential applications in the 
design of low cost, highly efficient synchronization solutions. The finding 
that synchronization can be improved by partial phase modulation provides 
useful insights into the control of dynamics in complex nonlinear systems - 
a challenging problem in nonlinear science and complex systems at present. 

That synchronization performance can be improved by introducing 
phase lags to a fraction of all oscillators has implications to the problem
of controlling complex nonlinear networks. In an early study, phase 
modulation was proposed for harnessing chaos in low-dimensional nonlinear 
systems, e.g., the chaotic Duffing oscillator~\cite{QHYQ:1995}. What we 
have achieved here is to extend the idea of phase control to optimizing 
synchronization in complex spatiotemporal systems consisting of an 
ensemble of oscillators. As phase lags are ubiquitous and can be modulated
in realistic systems, we hope that the strategy will find applications in
controlling and optimizing complex infrastructure and biological systems, 
e.g., the power grid and neuronal networks.

\section*{Acknowledgement}

This work was supported by the National Natural Science Foundation of 
China under the Grant No.~11375109 and by the Fundamental Research Funds 
for the Central Universities under the Grant Nos.~GK201601001 and 2017TS003. 
YCL was supported by ONR through Grant No.~N00014-16-1-2828.

\end{document}